\documentclass[12pt]{iopart}

\usepackage{iopams}
\usepackage{graphicx}
  
\begin{document}

\title{Non-Hermitian degeneracy of two unbound states} \author{E.
  Hern\'andez$^{1}$, A. J\'auregui$^{2}$ and A. Mondrag\'on$^{1}$ }

\address{$^{1}$Instituto de F\'{\i}sica, UNAM, Apdo. Postal 20-364,
01000 M\'exico  D.F. M\'exico}
\address{$^{2}$Departamento de F\'{\i}sica, Universidad de Sonora,
  Apdo. Postal 1626, Hermosillo, Sonora, M\'exico}
\ead{mondra@fisica.unam.mx}

\begin{abstract}
 We solved numerically the implicit, trascendental equation that
 defines the eigenenergy surface of a degenerating isolated doublet of
 unbound states in the simple but illustrative case of the scattering
 of a beam of particles by a double barrier potential. Unfolding the
 degeneracy point with the help of a contact equivalent approximant,
 crossings and anticrossings of energies and widths, as well as the
 changes of identity of the poles of the $S-$matrix are explained in
 terms of sections of the eigenenergy surfaces.

\end{abstract}


\maketitle

\section{Introduction}

Unbound decaying states are energy eigenstates of a time reversal
invariant Hamiltonian describing non-dissipative physics in a
situation in which there are no particles incident \cite{one,two}. Even when
the formal Hamiltonian, considered as an operator in the Hilbert space
of square integrable functions, is Hermitian (self-adjoint), this
boundary condition makes the eigenvalue problem non-selfadjoint and the
corresponding energy eigenvalues complex, ${\cal E}_{n} = E_{n} - i
1/2\Gamma_{n}$ with $E_{n} > \Gamma_{n} > 0$ \cite{three}.

Commonly, unbound energy eigenstates are regarded as arising from a
perturbation with the physics essentially unchanged from the bound
state case, except for an exponential decay. But, unbound state
physics differs radically from bound state physics in the presence of
degeneracies that is, coalescence of eigenvalues \cite{four}.

In the case of a Hermitian Hamiltonian depending on parameters, the
bound state energy eigenvalues are real and, when a single parameter
is varied, the mixing of two levels leads to the well known phenomenon
of energy level repulsion \cite{five} and avoided level crossing
\cite{six}. In the case of unbound energy eigenstates (resonances) of
the same Hamiltonian, the energy eigenvalues are complex and, when a
single parameter is varied, this fact opens a rich variety of
possibilities, namely, crossings and anticrossings of energies and
widths \cite{seven}, and the so called ``change of identity'' of the
poles of the $S-$matrix \cite{eight}. These novel effects have
attracted considerable theoretical \cite{nine,ten,eleven} and recently
also experimental interest \cite{twelve}. True degeneracies of
resonance energy eigenvalues result from a joint crossing of energies
and widths in a physical system depending on only two real parameters
and give rise to the occurrence of a double pole of the scattering
matrix in the complex energy plane \cite{ten}.  Associated with a
double pole of the $S-$matrix, the quantum system has a chain of
Jordan-Gamow generalized eigenfunctions \cite{three,thirten,fourteen}.
Examples of double poles in the scattering matrix of simple quantum
systems have been recently described
\cite{fifteen,sixteen,seventeen,eighteen}. Degeneracy of complex
energy eigenvalues of non-Hermitian PT symmetric Hamiltonian has been
discussed by M. Znojil \cite{znojil,znojil1,znojil2} and A. Mostafazadeh
\cite{mostafa}.  From a phenomenological point of view, I. Rotter
discussed double poles in the scattering matrix and level repulsion of
unbound states in the framework of an effective many body theory
\cite{rotter1,rotter2,rotter3,rotter4} .

The characterization of the singularities of the eigenenergy surfaces
at a degeneracy of unbound states in parameter space arises naturally
in connection with the Berry phase of unbound states that was
predicted by Hern\'andez et al \cite{nineteen,twenty,twentyone} and
independently by W.D. Heiss \cite{twentytwo}, see also the interesting
recent work by A.A. Mailybaev et al \cite{alexei}.  The Berry phase of
two unbound states was measured by the Darmstadt group
\cite{twentythree,twentyfour}. The unfolding of energy eigenvalue
surfaces at a degeneracy of unbound states of a Hermitian Hamiltonian
was discussed by E. Hern\'andez et al \cite{twentyfive} and the
unfolding of eigenvalue surfaces of non-Hermitian Hamiltonian matrices
with applications in modern problems of quantum physics, cristal
optics, physical chemistry, acoustics, mechanics and circuit theory
has been the subject of many recent investigacions
\cite{twentysix,twentyseven,twentyeight,twentynine,thirty,thirtyone,
  heiss2,heiss3}.

In this paper, we will be concerned with some physical and mathematical
aspects of the mixing and degeneracy of two unbound energy eigenstates
in an isolated doublet of resonances of a quantum system depending on
two control parameters. The plan of the paper is as follows: In
section 2, we give the results of a numerical computation of the
surfaces that represent the resonance energy eigenvalues as functions
of the control parameters in the scattering of a beam of particles by
a double barrier potential. The analytical structure of the singularity
of the energy surface at the crossing point is characterized in
section 3, where we also introduce a contact equivalent approximant to
the energy surface at the degeneracy point. Section 4 is devoted to a
discussion of crossings and anticrossings of energies and widths, as
well as, the changes of identity of the poles of the $S-$matrix, in
term of sections of the energy surfaces. We end our paper with a short
summary and some conclusions.

\section{Resonances in  a double barrier potential}
Doublets of resonances and accidental degeneracies of unbound states
may occur in the scattering of a beam of particles by a potential with
two regions of trapping. A simple example is provided by a spherically
symmetric potential $V(r)$ such that the two regions of trapping are
two potential wells defined by two concentric potential barriers
located between the origin of coordinates and the outer region where
the potential $V(r)$ vanishes. In order to make the analysis as simple
and explicit as possible, we take the wells and barriers to be square
as shown in figure 1.
 \begin{figure}
\begin{center}
\includegraphics[width=220pt,height=150pt]{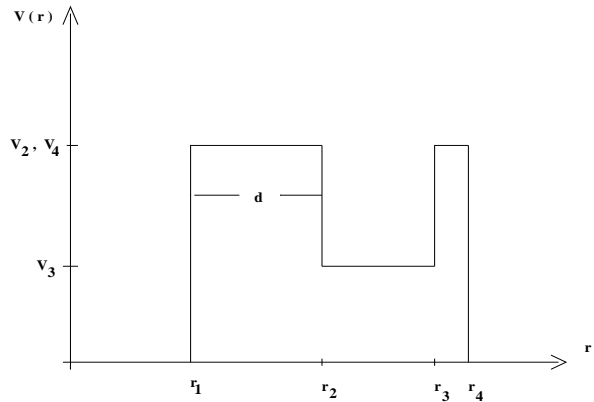}
\caption{The double barrier potential is such that it supports two
  unbound states with equal energies and half-lives. The control
  parameters of the system are $d$ and $V_{3}$.}
\end{center}
\end{figure}
In this section, we will consider the conditions for the occurrence of
a  degeneracy of unbound  states in   this simple  system and  we will
describe the numerically     exact computation of the  surfaces   that
represent  the complex energy eigenvalues as  functions of the control
parameters of the  system in the  neighbourhood of and at a degeneracy
of unbound states.

\subsection{The  Jost regular solution}
The s-wave radial Schr\"odinger equation is
\begin{eqnarray}\label{uno}
\frac{d^{2}\phi(k,r)}{dr^{2}} + (k^{2} - U(r) ) \phi(k,r) = 0,
\end{eqnarray}
the potential $U(r)= 2m V(r)/\hbar^{2}$ is a double barrier such that
between the origin of coordinates and the outer region,
where the particles propagate freely, there are two square potential
wells separated by two square potential barriers, as shown in figure 1.
The system has seven parameters, the positions $r_{i}$ (i = 1,2,3,4)
and heights $V_{i}$ (i = 2,3,4) of the four discontinuities of the
potential. We will keep the five parameters $(
V_{2}, V_{4}, r_{1}, r_{3} - r_{2}, r_{4} - r_{3})$ fixed and will
vary the depth $V_{3}$ of the outer well  and the thickness of the
inner barrier $d = r_{2} - r_{1}$. In the following, we will refer to
the pair of parameters $(d,V_{3})$ as the control parameters of the
system.

The radial Schr\"odinger equation (\ref{uno}) is solved exactly.
The Jost regular solution of (\ref{uno}) normalized to unit slope at the
origen, $\phi (k,r)$, is as follows:

In the wells,
\begin{eqnarray}\label{dos}
\phi_{1}(k,r) = \frac{1}{k} \sin k r, \hspace{1.5cm} 0\leq r \leq r_{1},
\end{eqnarray}
and
\begin{eqnarray}\label{tres}
\phi_{3}(k,r)  &=& \phi_{2} (k,r_{2})\Bigl[\cos \Bigl(K_{3}(k, V_{3})(r-r_{2})
\Bigr) \nonumber \\ 
&+& \alpha_{2} (k,d)\sin \Bigl(K_{3}(k, V_{3})(r-r_{2})\Bigr) \Bigr],
 \hspace{0.3cm}
r_{2}\leq r \leq r_{3}.
\end{eqnarray}
In the barriers,
\begin{eqnarray}\label{cuatro}
\phi_{i}(k,r) &=& \phi_{i-1}(k,r_{i-1})\Bigl[\cosh
\Bigl(K_{i}(k)(r-r_{i-1})\Bigr) \nonumber \\
&+& \alpha_{i-1}(k,d)
\sinh\Bigl(K_{i}(k)(r-r_{i-1})\Bigr)\Bigr],  
\hspace{0.5cm}  r_{i-1}\leq
r\leq r_{i}, \ \ \ i = 2,4 , \nonumber \\
\end{eqnarray}
and, in the outer region,
\begin{eqnarray}\label{cinco}
\phi_{5}(k,r) = \phi_{4}(k,r_{4})\Bigl[\cos k(r-r_{4}) +
\alpha_{4}(k,d)\sin k(r-r_{4})\Bigr], \ \ \
 r_{4}  \leq r < \infty. \nonumber \\
\end{eqnarray}

In these expressions $k$ is the wave number of the free waves and
\begin{eqnarray}\label{seis}
K_{i}(k) = \Bigl( (-1)^{i}( U_{i} - k^{2}) \Bigr)^{1/2}, \hspace{1cm}i =
2,3, 4
\end{eqnarray}
is the wave number in the barriers and the outer well.

The logaritmic derivatives $\alpha_{i}(k,d, V_{3})$ of $\phi(k,r)$ at
the consecutive discontinuities $r_{i}$ and $r_{i+1}$ are related by
the matching conditions at $r_{i+1}$,
\begin{eqnarray}\label{siete}
\alpha_{1}(k) = \frac{k}{K_{2}(k)}\cot kr_{1},
\hspace{0.7cm}
\alpha_{2}(k,d) = \frac{K_{2}(k)}{K_{3}(k)}\frac{\alpha_{1}(k)+\tanh
  (K_{2}(k)d)}{1+\alpha_{1}(k)\tanh (K_{2}(k)d)},
\end{eqnarray}
\begin{eqnarray}\label{ocho}
\alpha_{3}(k;d,V_{3}) = \frac{K_{3}(k,V_{3})}{K_{4}(k)}\frac{\alpha_{2}(k,d) -
  \tan (K_{3}(k)(r_{3} - r_{2}) )}{1 + \alpha_{2}(k,d)\tan(K_{3}
  (k)(r_{3} - r_{2} ))}, 
\end{eqnarray}
and
\begin{eqnarray}\label{nueve}
\alpha_{4}(k;d,V_{3}) = \frac{K_{4}(k)}{k}\frac{\alpha_{3}(k;d,V_{3}) + \tanh
  (K_{4}(k)(r_{4} - r_{3}) )}{1 + \alpha_{3}(k;d,V_{3})\tanh (K_{4}(k)(r_{4}
  - r_{3}) )}. 
\end{eqnarray}

Since the first logaritmic derivative is known, succesive substitution
of $\alpha_{i}(k,d, V_{3})$ in $\alpha_{i+1}(k,d,V_{3})$ gives an
explicit solution for $\alpha_{4}(k;d,V_{3})$ and an explicit
expression for the regular solution is obtained from
equations (\ref{dos}-\ref{cinco}).

\subsection{The Jost function}
When the regular solution in the outer region, (\ref{cinco}), is
written as a linear combination of an outgoing wave $\exp(ikr)$ and an
incoming wave $\exp{(-ikr)}$
\begin{eqnarray}\label{diez}
\phi_{5}(k,r) &=& \phi_{4}(k,r_{4})\frac{1}{2}\Bigl[\Bigl(1 -
i\alpha_{4}(k;V_{3},d) \Bigr)\exp {ik(r-r_{4})} \cr 
&+& \Bigl(1 + i\alpha_{4}(k;V_{3},d) \Bigr)\exp {-ik(r-r_{4})} \Bigr], 
 \hspace{1cm} r_{4} \leq r < \infty, 
\end{eqnarray}
the coefficient of the incoming wave is the Jost function. Making use
of equations (\ref{cuatro}), (\ref{nueve}) and (\ref{diez}) we find an exact
expression for the Jost function $f(-k)$,
\begin{eqnarray}\label{once}
f(-k; d, V_{3}) = \sin kr_{1}\Bigl[\cosh K_{2}(k)d  +  
\alpha_{1}(k)
\sinh K_{2}(k)d\Bigr] \nonumber \\
\times \Bigl[\cos K_{3}(k, V_{3})(r_{3} -r_{2})  
+ \alpha_{2}(k,d, V_{3})\sin K_{3}(k, V_{3})
(r_{3} - r_{2})\Bigr] \nonumber \\
\times  \Bigl\{\frac{K_{4}(k)}{k} \Bigl[\sinh
K_{4}(k)(r_{4}-r_{3})  
+ \alpha_{3}(k;d,V_{3})\cosh K_{4}(k)(r_{4}-r_{3})\Bigr] 
\nonumber  \\ 
 - i\Bigl[\cosh K_{4}(k)(r_{4} - r_{3})  
+ \alpha_{3}(k;d,V_{3})\sinh K_{4}(k)(r_{4}-r_{3})\Bigr]\Bigr\}
\exp ikr_{4} 
\end{eqnarray}

\subsection{The physical solutions}
The scattering wave function, $\psi(k,r)$, and the regular
solution $\phi (k,r)$ are related by \cite{thirtytwo}
\begin{eqnarray}\label{doce}
\psi(k,r) = \frac{-2ik}{f(-k)}\phi(k,r),
\end{eqnarray}
and the scattering matrix is given by
\begin{eqnarray}\label{trece}
S(k) = \frac{f^{*}(-k)}{f(-k)} = \exp (i2\delta(k) ),
\end{eqnarray}
where the Jost function $f(-k)$ is given by (\ref{once}).

The zeros of the Jost function give resonance poles in the scattering
wave function $\psi(k,r)$, and in the $S(k)$ matrix, and from (\ref{uno})
and (\ref{dos}-\ref{cinco}), we may verify that all unbound energy
eigenfunctions of the radial Schr\"odinger equation are associated
with roots (zeros) of the Jost function.

Unbound state eigenfunctions also called resonant-state or Gamow
eigenfunctions are the solutions of equation (\ref{uno}) that vanish
at the origin, and at infinity satisfy the outgoing wave boundary
condition.  When the Jost function has a zero at $k_{n}$, the
coefficient of the incoming wave in (\ref{diez}) vanishes, and
$\phi(k_{n},r)$ is proportional to the outgoing wave solution of
equation (\ref{uno}) for $r$ larger than the range of the potential.
Hence, the unbound state eigenfunctions are related to the regular
solution by
\begin{eqnarray}\label{catorce}
u_{n}(k_{n},r) = N^{-1}_{n}\phi(k_{n},r),
\end{eqnarray}
where $N_{n}$ is a normalization constant and $k_{n}$ is a zero of the
Jost function
\begin{eqnarray}\label{quince}
f(-k_{n};d,V_{3}) = 0
\end{eqnarray}

\subsection{Degeneracy of unbound energy eigenvalues}
A degeneracy of unbound states results from the exact coincidence of
two simple zeros of the Jost function, which merge into one double
zero lying in the fourth quadrant of the complex $k-$plane. Hence, the
condition for  a degeneracy of two unbound energy
eigenstates at some $k = k_{d}$ is that both, the Jost function and
its first derivative vanish at $k_{d}$,
\begin{eqnarray}\label{dieciseis}
f(-k_{d}; d, V_{3}) = 0,
\hspace{1.3cm}
\Bigl(\frac{df(-k; d, V_{3})}{dk}\Bigr)_{k_{d}} = 0,
\end{eqnarray}
where $f(-k; d, V_{3})$ is given in (\ref{once}).

Therefore, to locate a degeneracy of unbound states, the coupled
equations (\ref{dieciseis}) were solved numerically. The zeroes of the
Jost function are found by an algebraic computer package that searches
for the minima of $|f(-k)|$ in the complex k-plane. In the computation
the five parameters $V_{2}, V_{4}, r_{1}, r_{3}-r_{2}$ and
$r_{4}-r_{3}$ were kept fixed at the values $V_{2}=V_{4}=2, r_{1} = 1,
r_{3}-r_{2} = 1, r_{4}-r_{3} =$ 0.304892 and only the control
parameters $d$ and $V_{3}$ were allowed to vary. Starting with the
values $d=$ 2 and $V_{3}=$ 1.04, we find the first isolated doublet of
resonances at $k_{1} = $ 2.2101546 - i 0.1366887 and $k_{2} =$
2.2321776 - i 0.0017984.

Then, we adjusted the control parameters $d$ and $V_{3}$ until 
$k_{1}$ and $k_{2}$ became equal to some common
value $k_{d}$. We also  computed numerically $|df(-k)/dk|$ at $k =
k_{d}$ to verify that the second equation was also satisfied to
some previously prescribed accuracy. In this way, we found
that by fine tuning the control parameters to the values $d^{*} =
1.1314661145$ and $V_{3}^{*} = 1.038235081$, the first doublet of
resonances becomes degenerate, with a precision better than one part
in $10^{8}$, at $k_{d} =$ 2.22697606 - i 0.07220139.

In the following we will refer to the double zero of $f(-k)$ at
$k_{d}$ as the degeneracy point or crossing point of the doublet of
resonant states in the complex k-plane and to the point
$(d^{*},V^{*}_{3})$ as the exceptional point in parameter space.

\subsection{Energy surfaces}
The energy eigenvalues ${\cal E}_{n}(d,V_{3}) =
(\hbar^{2}/2m)k^{2}_{n}(d,V_{3})$ of the physical system are obtained
from the zeroes of the Jost function, given in equation (\ref{quince}). 
That condition defines, implicitly, the inverse functions
\begin{eqnarray}\label{diecinueve}
k_{n}(d,V_{3}) = f^{-1}(0; d, V_{3}),\hspace{1cm} n = 1,2,....
\end{eqnarray}
as branches of a multivalued function \cite{thirtytwo} which will be
called the wave number pole position function. Each branch
$k_{n}(d,V_{3})$ is a continuous, single valued function of the
control parameters.  When the physical system has an isolated doublet
of unbound states which become degenerate for some exceptional values
of the control parameters $(d^{*},V_{3}^{*})$, the corresponding two
branches of the energy pole position function, ${\cal E}_{1}(d,V_{3})$
and ${\cal E}_{2}(d,V_{3})$, are equal (cross or coincide) at that
point.
\begin{figure}
\begin{center}
\includegraphics[width=190pt,height=200pt]{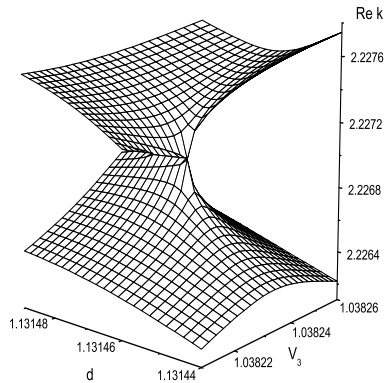}
\caption{The two - sheeted surface $S_{R}$ that represents the real
  part of the eigenwave numbers $k_{1}$ and $k_{2}$ as functions of
  the control parameters $(d,V_{3})$ in the vecinity of a degeneracy
  of unbound states. }
\end{center}
\end{figure}
With the purpose of exploring the geometrical and topological
properties of the surfaces representing the pole position function
$k_{1,2}(d,V_{3})$ of the isolated doublet of resonances in parameter
space, we solved numerically the implicit equation (\ref{quince}) for
$k_{1}(d,V_{3})$ and $k_{2}(d,V_{3})$ in the neighbourhood of and at a
degeneracy of unbound states. The results of the numerical computation
are represented as surfaces in a Euclidean space with coordinates (Re
$k$, Im $k$, d, $V_{3}$).

In figure 2,  the real function Re $k_{1,2}(d,V_{3})$ is shown as a
surface $S_{R}$ in the three-dimensional subspace with cartesian
coordinates (Re $k, d, V_{3}$). Similarly, in figure 3, the real
function Im $k_{1,2}(d, V_{3})$ is shown as a surface $S_{I}$ in the
three-dimensional subspace with cartesian coordinates (Im $k, d,
V_{3}$).

We see that close to the degeneracy of unbound states, the function Re
$k_{1,2}(d, V_{3}$) has two branches and is represented by a two
sheeted surface $S_{R}$. The two sheets of $S_{R}$ are two copies of
the plane $(d,V_{3})$ which are cut and joined smoothly along a line
${\cal L}_{R}$ starting at the exceptional point and extending to
values $d\geq d^{*}$ and $V_{3} \geq V^{*}_{3}$.
\begin{figure}
\begin{center}
\includegraphics[width=190pt,height=200pt]{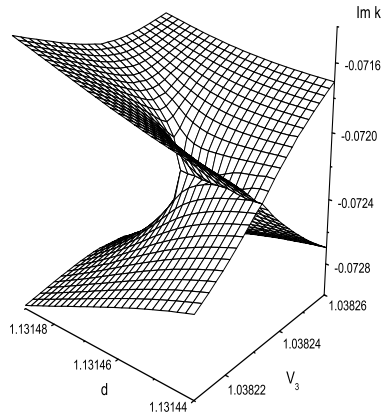}
\caption{The two-sheeted surface $S_{I}$ that represents the imaginary
  part of eigenwave numbers $k_{1}$ and $k_{2}$ as functions of the
  control parameters $(d,V_{3})$ in the neighbourhood of a degeneracy
  of unbound states with complex resonance energies ${\cal E}_{i} =
  \hbar^{2}k^{2}_{i}/2m$, $i = 1,2$. }
\end{center}
\end{figure}

The function Im $k_{1,2}(d, V_{3})$ has also two branches and is
represented by a two sheeted surface $S_{I}$. The two sheets of $S_{I}$
are two copies of the plane $(d, V_{3})$ which are cut and pasted
smoothly along a line ${\cal L}_{I}$ extending from the exceptional
point to values $d \leq d^{*}$ and $V_{3} \leq V^{*}_{3}$.

The projection of the lines ${\cal L}_{R}$ and ${\cal L}_{I}$ on the
plane $(d, V_{3})$ are the two halves of a line ${\cal L}'$ that goes
through the exceptional point $(d^{*}, V^{*}_{3})$.

At the exceptional point, and only at that point, both the real and
imaginary parts of $k_{1}$ and $k_{2}$ are equal. Therefore, in the
complex k-plane, at the crossing point, the two simple zeroes of the
Jost function merge into one double zero which is an isolated point in
the complex k-plane.

\section{The analytical behaviour of the pole-position function close
  to the exceptional point}
 In this section, it will be shown that, the singularity of the energy
surface at the degeneracy or crossing point is an algebraic branch
point of square root type.  The analytical structure of the
singularity of the energy surface will be worked out and discussed in
detail. The results in this and the following section are not
restricted to the double barrier potential, to stress this generality,
the control parameters will be called $x_{1}$ and $x_{2}$.

Let us start by recalling that, when the first and second absolute
moments of the potential exist, and the potential decreases at
infinity faster than any exponential [e.g., if $V(r)$ has a Gaussian
tail or if it vanishes identically beyond a finite radius], the Jost
function $f(-k; x_{1},x_{2})$, is an entire function of $k$
\cite{thirtytwo}. The entire function of $k$, $f(-k;x_{1},x_{2})$, may be
written in the form of an infinite product, according to Hadamard's
form of the Weierstrass factorization theorem \cite{thirtythree}, and
by using a theorem of Pfluger \cite{thirtyfour}, we may write
\begin{eqnarray}\label{jos}
f(-k;x_{1},x_{2})= f(0)\exp({ikR})\prod_{1}^{\infty}\Bigl(1-
\frac{k}{k_{n}(x_{1},x_{2})}\Bigr)
\end{eqnarray}
$R$ is the range of the potential, $f(0)\neq 0$ and $\{k_{n}\}$ are the
zeroes of $f(-k;x_{1},x_{2})$, see also R.G. Newton \cite{thirtytwo}.

When the Jost function has a set of isolated zeroes, the implicit
equation for the pole position function, equation (\ref{quince}), may,
in principle, be solved, at least numerically, for each branch
$k_{n}(x_{1},x_{2})$ without ambiguity. When the system has an
isolated doublet of resonances which may become degenerate, the
corresponding branches of the pole position function, say
$k_{1}(x_{1},x_{2})$ and $k_{2}(x_{1},x_{2})$, may be equal (cross or
coincide) at an exceptional point. In this case, it is not possible to
solve equation (\ref{quince}) for each individual branch without
ambiguity and one should proceed to solve that equation for the pole
position function of the two members of the isolated doublet of
resonances.

 To be precise, we will say that the system has an isolated
doublet of unbound states if there is a finite bounded and connected
region ${\cal M}$ in parameter space and a finite domain ${\cal D}$ in
the fourth quadrant of the complex $k-$plane, such that when
$(x_{1},x_{2})\ \epsilon \ {\cal M}$, the Jost function has two and only two
zeroes, $k_{1}$ and $k_{2}$, in the finite domain ${\cal D} \subset {\cal
  C}$, all other zeroes of $f(-k; x_{1}, x_{2})$ lying outside ${\cal
  D}$.

The pole position function $k_{1,2}(x_{1},x_{2})$ of the isolated
doublet of resonances,
\begin{eqnarray}\label{veintiuno}
k_{1,2}(x_{1},x_{2}) = f^{-1}(0; x_{1},x_{2}),
\end{eqnarray}
is implicitly defined by the equation $f(-k_{1,2};x_{1},x_{2}) = 0$, and the
conditions
\begin{eqnarray}\label{veintidos}
\Bigl(\frac{df(-k;x_{1},x_{2})}{dk}\Bigr)_{k_{d}} = 0, 
\hspace{1cm}\mbox{and} \hspace{1cm}
\Bigl(\frac{d^{2}f(-k;x_{1},x_{2})}{dk^{2}}\Bigr)_{k_{d}} \neq 0,
\end{eqnarray}
for $(x_{1},x_{2})$ in a neighbourhood of the exceptional point
$(x_{1}^{*},x_{2}^{*})$.

 Then, to find an expression for the pole position function of
the isolated doublet of unbound states, the two zeroes of
$f(-k;x_{1},x_{2})$, corresponding to the isolated doublet of unbound
states are explicitly factorized in (\ref{jos}) as
\begin{eqnarray}\label{jo0}
f(-k;x_{1},x_{2}) = \Bigl[k - k_{1}(x_{1},x_{2})\Bigr]
\Bigl[k-k_{2}(x_{1},x_{2})\Bigr]g_{1,2}(k;x_{1},x_{2}),
\end{eqnarray}
which may be conveniently rearranged as
\begin{equation}\label{veinte}
\begin{array}{c}
f(-k; x_{1},x_{2}) = \Bigl[\bigl(k - \frac{1}{2}(k_{1}+k_{2})\bigr)^{2}
-  \frac{1}{4}\bigl(k_{1} - 
k_{2}\bigr)^{2}\Bigr] 
g_{1,2}(k,x_{1},x_{2}).
\end{array}
\end{equation}
with
\begin{eqnarray}\label{g12}
g_{1,2}(k;x_{1},x_{2})= f(0)\exp(ikR)\frac{1}{k_{1}(x_{1},
x_{2})k_{2}(x_{1},x_{2})}\prod_{3}^{\infty}\Bigl(1-\frac{k}{k_{n}(x_{1},x_{2})}\Bigr),
\cr
\end{eqnarray}
the expression in square brackets in the right hand side of equation
(\ref{veinte}) is the  Weierstrass polynomial of the isolated
doublet of unbound states\cite{thirtyfive}.

Solving equation (\ref{veinte}) for $k_{1,2}(x_{1},x_{2})$ when
$f(-k;x_{1},x_{2})$ vanishes, we get

\begin{eqnarray}\label{veintitres}
k_{1,2}(x_{1},x_{2}) = \frac{1}{2}\Big(k_{1}(x_{1},x_{2}) +
    k_{2}(x_{1},x_{2})\Bigr)  + 
    \sqrt{\frac{1}{4}\Big(k_{1}(x_{1},x_{2}\bigr)
      -k_{2}\bigl(x_{1},x_{2}\bigr)\Bigr)^{2}} \cr 
\end{eqnarray}
with $(x_{1},x_{2})$ in a neighbourhood of the exceptional point. This
equation relates the wave number pole position function of the doublet
to the pole position function of the individual unbound (resonance)
states.  Since the argument of the square-root function is complex, it
is necessary to specify the branch. Here and thereafter, the square
root of any complex quantity $F$ will be defined by
\begin{eqnarray}\label{veinticuatro}
\sqrt{F} = |\sqrt{F}| \exp\bigl(i\frac{1}{2}arg F\bigr), \hspace{1cm}
0\leq arg \ F \leq 2\pi
\end{eqnarray}
so that $|\sqrt{F}|= \sqrt{|F|}$ and the $F -$ plane is cut along the
real axis.

Now, we will proceed to the derivation of a contact equivalent
approximant to the pole position function of the doublet at the
crossing point.

According to the preparation theorem of Weierstrass \cite{thirtyfive},
the functions $1/2(k_{1}(x_{1},x_{2})+k_{2}(x_{1},x_{2}))$ and
$1/4(k_{1}(x_{1},x_{2})-k_{2}(x_{1},x_{2}))^{2}$, appearing in the
right hand side of equation (\ref{veintitres}), are regular at the
exceptional point and admit a Taylor series expansion about that
point,
\begin{eqnarray}\label{veinticinco}
\frac{1}{2}\Bigl(k_{1}(x_{1},x_{2}) + k_{2}(x_{1},x_{2})\Bigr) &=&
k_{d}(x^{*}_{1},x^{*}_{2}) \cr &+& 
\sum^{2}_{i=1}d^{(1)}_{i}(x_{i}-x^{*}_{i}) +
O\bigl((x_{i}-x^{*}_{i})^{2}\bigr)
\end{eqnarray}
and
\begin{equation}\label{veintiseis}
\begin{array}{c}
\Bigl(k_{1}(x_{1},x_{2}) - k_{2}(x_{1},x_{2})\Bigr)^{2} =
 \sum^{2}_{i=1}c^{(1)}_{i}(x_{i} - x^{*}_{i}) +
O\bigl((x_{i}-x_{i}^{*})^{2}\bigr).
\end{array}
\end{equation}

The complex coefficients $c^{(1)}_{i}$ and $d^{(1)}_{i}$, appearing in
these equations, may readily be computed from the Jost function with
the help of the implicit function theorem \cite{thirtyfive},
\begin{equation}\label{ventisiete}
\begin{array}{c}
c^{(1)}_{i} =  
\frac{-8}{\Bigl[\Bigl(\frac{\partial^{2}f(-k;x_{1},x_{2})}{\partial
    k^{2}}\Bigr)_{x^{*}_{1},x^{*}_{2}}\Bigr]_{k=k_{d}} }\Bigl[\Bigl
(\frac{\partial f(-k;x_{1},x_{2})}{\partial
  x_{1}}\Bigr)_{x_{2}}\Bigr]_{k=k_{d}}, 
\end{array}
\end{equation}
and
\begin{equation}\label{veintiocho}
\begin{array}{c}
d^{(1)}_{i} = 
\frac{-1}{\Bigl[\Bigl(\frac{\partial^{2}f(-k;x_{1},x_{2})}{\partial
    k^{2}}\Bigr)_{x^{*}_{1},x^{*}_{2}}\Bigr]_{k=d_{d}}} 
\Bigl\{\Bigl[\Bigl(\frac{\partial^{2}f(-k;x_{1},x_{2})}{\partial
  x_{1}\partial k}\Bigr)_{x_{2}}\Bigr]_{k=k_{d}} - \cr
\frac{1}{\Bigl[\Bigl(\frac{\partial^{2}f(-k;x_{1},x_{2})}{\partial
    k^{2}}\Bigr)_{x^{*}_{1},x^{*}_{2}}\Bigr]_{k=k_{d}} } 
  \frac{1}{3}\Bigl[\Bigl(\frac{\partial^{3}f(-k;x_{1},x_{2})}{\partial
  k^{3}}\Bigr)_{x^{*}_{1},x^{*}_{2}}\Bigr]_{k=k_{d}} \cr \times \Bigl[\Bigl(
\frac{\partial f(-k;x_{1},x_{2})}{\partial x_{1} }\Bigr)_{x_{2}}
\Bigr]_{k=k_{d}}\Bigr\}.
\end{array}
\end{equation}
A contact equivalent approximant, $\hat{k}_{1,2}(x_{1},x_{2})$, to the
doublet's pole position function is obtained when the Taylor series
expansions (\ref{veinticinco}) and (\ref{veintiseis}) are
substituted for the functions
$1/2(k_{1}(x_{1},x_{2})+k_{2}(x_{1},x_{2}))$ and
$1/4(k_{1}(x_{1},x_{2})-k_{2}(x_{1},x_{2}))^{2}$. Keeping only the
first order terms, we obtain
\begin{eqnarray}\label{veintinueve}
\hat{k}_{1,2}(x_{1},x_{2}) &=& k_{d}(x^{*}_{1},x^{*}_{2}) + \sum_{i=1}^{2}
d^{(1)}_{i}(x_{i}-x_{i}^{*}) \cr &+& 
\sqrt{\frac{1}{4}\bigl[c^{(1)}_{1}(x_{1}-x^{*}_{1}) 
+  c^{(1)}_{2}(x_{2}-x^{*}_{2})\bigr] }.
\end{eqnarray}

Then, from (\ref{veinte}) and (\ref{veintinueve}), close to the
  exceptional point, the Jost function may be approximated as
\begin{equation}\label{treinta}
f(-k; x_{1},x_{2}) \approx \frac{1}{g_{1,2}(k_{d};x^{*}_{1},x^{*}_{2})}
\hat{f}_{doub}(-k;x_{1},x_{2}),
\end{equation}
where
\begin{eqnarray}\label{treintaiuno}
\hat{f}_{doub}(-k;x_{1},x_{2}) &=& \Bigl[k -\Bigl(k_{d} +
\sum^{2}_{i=1}d^{(1)}_{i}(x_{i},x_{i}^{*})\Bigr)\Bigr]^{2} \cr
&-& \frac{1}{4}\Bigl(c^{(1)}_{1}(x_{1}-x^{*}_{1}) + 
c^{(1)}_{2}(x_{2}-x^{*}_{2}) \Bigr).
\end{eqnarray}

The coefficient $\bigl[g_{1,2}(k_{d},x_{1}^{*},x^{*}_{2})\bigr]^{-1}$
in (\ref{treinta}) may be understood as a scaling factor. Hence, the
two parameter family of functions $\hat{f}_{1,2}(-k;x_{1},x_{2})$ is
contact equivalent to the Jost function at the exceptional point and
is also a universal unfolding \cite{thirtysix} of $f(-k;x_{1},x_{2})$
at the exceptional point where the degeneracy of unbound states occurs.

\subsection{Energy-pole position function}
A contact equivalent approximant to the energy pole position function
${\cal E}_{1,2}(x_{1},x_{2})$ at the crossing point of the doublet of
unbound states is readily obtained from
(\ref{veintitres},\ref{veinticinco}-\ref{veintiseis}). Taking
the square in both sides of (\ref{veintinueve}), multiplying them by
$\hbar^{2}/2m$ and recalling ${\cal E}_{i} = (\hbar^{2}/2m)k^{2}_{i}$,
in the approximation of (\ref{veinticinco}-\ref{veintinueve}), we
get
\begin{eqnarray}\label{treintaidos}
{\cal E}_{1,2}(x_{1},x_{2}) \approx {\cal
  E}_{d}(x^{*}_{1},x^{*}_{2}) +
\Delta {\cal E}_{d}(x_{1},x_{2})  + \hat{\epsilon}_{1,2}(x_{1},x_{2}),
\end{eqnarray}
where
\begin{eqnarray}\label{treintaitres}
\hat{\epsilon}_{1,2}(x_{1},x_{2}) =
\sqrt{\frac{1}{4}\bigl[C^{(1)}_{1}(x_{1} - x^{*}_{1}) +
  C^{(1)}_{2}(x_{2}-x_{2}^{*})\bigr]},
\end{eqnarray}
and $C^{(1)}_{i} = \Bigl(\hbar^{2}k_{d}(x^{*}_{1},x^{*}_{2})/m
\Bigr)^{2}c^{(1)}_{i}$. It will be convenient to change slightly the
notation
\begin{eqnarray}\label{treintaicuatro}
\vec{\xi} = \pmatrix{
\xi_{1} \cr
\xi_{2}
}
= \pmatrix{
x_{1} - x^{*}_{1} \cr
x_{2} - x^{*}_{2}
},
\hspace{0.5cm}\vec{R} = \pmatrix{
Re \ C^{(1)}_{1} \cr
Re \ C^{(1)}_{2}
},
\hspace{0.5cm}\vec{I} = \pmatrix{
Im \ C^{(1)}_{1} \cr
Im \ C^{(1)}_{2}
}.
\end{eqnarray}
The components of the real fixed vectors $\vec{R}$ and $\vec{I}$ are
the real and imaginary parts of the coefficients $C^{(1)}_{i}$ of
$(x_{i}-x^{*}_{i})$ in the Taylor expansion of the function
$1/4\bigl({\cal E}_{1}(x_{1},x_{2}) - {\cal
  E}_{2}(x_{1},x_{2})\bigr)^{2}$ and the real vector $\vec{\xi}$ is
the position vector of the point $(x_{1},x_{2})$ relative to the
exceptional point $(x^{*}_{1},x^{*}_{2})$ in parameter space.

In the notation defined in equations (\ref{treintaicuatro}),
\begin{eqnarray}\label{treintaicinco}
\hat{\epsilon}^{2}_{1,2}(x_{1},x_{2}) =
\frac{1}{4}\Bigl(\bigl(\vec{R}\cdot\vec{\xi}\bigr) +
i\bigl(\vec{I}\cdot\vec{\xi}\bigr)\Bigr)
\end{eqnarray}
and
\begin{eqnarray}\label{treintaiseis}
|\hat{\epsilon}_{1,2}(x_{1},x_{2})|^{2} =
+ \frac{1}{4}\sqrt{\Bigl(\bigl(\vec{R}\cdot\vec{\xi}\bigr)^{2} +
  \bigl(\vec{I}\cdot\vec{\xi}\bigr)^{2}\Bigr)}.
\end{eqnarray}

Solving for the real and imaginary parts of the function
$\hat{\epsilon}_{1,2}(x_{1},x_{2})$, we obtain
\begin{equation}\label{treintaisiete}
Re \ \hat{\epsilon}_{1,2}(x_{1},x_{2}) = \pm
\frac{1}{2\sqrt{2}}\Bigl[+\sqrt{\bigl(\vec{R}\cdot\vec{\xi}\bigr)^{2} +
  \bigl(\vec{I}\cdot\vec{\xi}\bigr)^{2}} + \vec{R}\cdot\vec{\xi}\Bigr]^{1/2}
\end{equation}
\begin{equation}\label{treintaiocho}
Im \ \hat{\epsilon}_{1,2}(x_{1},x_{2}) = \pm
\frac{1}{2\sqrt{2}}\Bigl[+\sqrt{\bigl(\vec{R}\cdot\vec{\xi}\bigr)^{2} +
  \bigl(\vec{I}\cdot\vec{\xi}\bigr)^{2}} - \vec{R}\cdot\vec{\xi}\Bigr]^{1/2}
\end{equation}
and
\begin{eqnarray}\label{treintainueve}
sgn \Bigl(Re \hat{\epsilon}_{1,2}\Bigr)sgn \Bigl(Im
\hat{\epsilon}_{1,2}\Bigr) = sgn \Bigl(\vec{I}\cdot \vec{\xi}\Bigr).
\end{eqnarray}
It follows from (\ref{treintaisiete}), that $Re \
\hat{\epsilon}_{1,2}(x_{1},x_{2})$ is a two branched function of
$(\xi_{1},\xi_{2})$ which may be represented as a two-sheeted surface
$S_{R}$ in a three dimensional Euclidean space with cartesian
coordinates $(Re {\cal E}, \xi_{1},\xi_{2})$. The two
branches of $Re {\cal E}(\xi_{1},\xi_{2})$ are represented
by two sheets which are copies of the plane $(\xi_{1},\xi_{2})$ cut
along a line where the two branches of the function are joined
smoothly. Since a negative and a positive numbers are equal only when
both vanish, the cut is defined as the locus of the points where the
argument of the square- root function in the right hand side of
(\ref{treintaisiete}) vanishes.  Close to the origen of coordinates (the
exceptional point), this locus is defined by a unit vector
$\hat{\xi}_{o}$ in the $(\vec{\xi}_{1},\vec{\xi}_{2})$, plane such
that
\begin{equation}\label{cuarenta}
\vec{I}\cdot\hat{\xi}_{o} = 0 \hspace{0.5cm}\mbox{and}\hspace{0.5cm}
\vec{R}\cdot\hat{\xi}_{o} = - |\vec{R}\cdot\hat{\xi}_{o}|
\end{equation}

Therefore, the real part of the energy-pole position function,
  ${\cal E}_{1,2}(x_{1},x_{2})$, as a function of the real
  parameters $(x_{1},x_{2})$, has an algebraic branch point of square
  root type (rank one) at the exceptional point with coordinates
  $(x^{*}_{1},x^{*}_{2})$ in parameter space, and a branch cut along a
  line, ${\cal L}_{R}$, that starts at the exceptional point and
  extends in the \underline{positive} direction defined by the unit
  vector $\hat{\xi}_{o}$ satisfying equations (\ref{cuarenta}).

A similar analysis shows that, the imaginary part of the
  energy-pole position function, $Im \ {\cal E}_{1,2}(x_{1},x_{2})$,
  as a function of the real parameters $(x_{1},x_{2})$, also has an
  algebraic branch point of square root type (rank one) at the
  exceptional point with coordinates $(x^{*}_{1},x^{*}_{2})$ in
  parameter space, and also has a branch cut along a line, ${\cal
    L}_{I}$, that starts at the exceptional point and extends in the
  \underline{negative} direction defined by the unit vector $\hat{\xi}_{o}$
  satisfying equations (\ref{cuarenta}).

The branch cut lines, ${\cal L}_{R}$ and ${\cal L}_{I}$, are in
orthogonal subspaces of a four dimensional Euclidean space with
coordinates $(Re {\cal E}, Im {\cal E}, \xi_{1},
\xi_{2})$ - but have one point in common, the exceptional point with
coordinates $(x^{*}_{1},x^{*}_{2})$.

Along the line ${\cal L}_{R}$, excluding the exceptional point
$(x^{*}_{1},x^{*}_{2})$,
Re ${\cal E}_{1}(x_{1},x_{2}) =$ Re $ {\cal E}_{2}(x_{1},x_{2})$
but Im $\ {\cal E}_{1}(x_{1},x_{2})
\neq $ Im $ {\cal E}_{2}(x_{1},x_{2})$.

Similarly, along the line ${\cal L}_{I}$, excluding the exceptional point,
Im $ {\cal E}_{1}(x_{1},x_{2}) = $ Im $ {\cal E}_{2}(x_{1},x_{2})$,
but Re $ {\cal E}_{1}(x_{1},x_{2})
\neq $ Re $ {\cal E}_{2}(x_{1},x_{2})$.

Equality of the complex resonance energy eigenvalues (degeneracy of
resonances), ${\cal E}_{1}(x_{1}^{*},x_{2}^{*}) = {\cal
  E}_{2}(x_{1}^{*},x_{2}^{*}) = {\cal E}_{d}(x^{*}_{1},x^{*}_{2})$,
occurs only at the exceptional point with coordinates
$(x^{*}_{1},x^{*}_{2})$ in parameter space and only at that point.

In consequence, in the complex energy plane, the crossing point of two
simple resonance poles of the scattering matrix is an isolated point
where the scattering matrix has one double resonance pole.

\section{Phenomenology of the exceptional point}
There is a variety of phenomenological manifestations of the
topological and geometrical properties of the singularity of the
energy surfaces at an exceptional point, which fall roughly in two
categories, according to the experimental set up. First, when one
control parameter is slowly varied while keeping the other constant,
cossings and anticrossings of energies and widths are experimentally
observed \cite{thirtyseven,thirtyeight,thirtynine}, as well as, the so
called, changes of identity of the poles of the scattering matrix
\cite{eight,fourty}.  Second, when the system is slowly
transported in a double circuit around the exceptional point in
parameter space, it is observed that the wave function of the system
acquires a geometrical or Berry phase
\cite{nineteen,twenty,twentyone,twentytwo,alexei,twentythree,twentyfour}.
\begin{figure}
\begin{center}
\includegraphics[width=220pt,height=230pt]{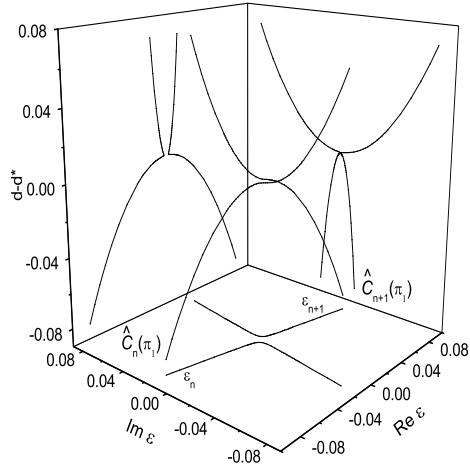}
\caption{The curves $\hat{C}_{n}(\pi_{1})$ and
  $\hat{C}_{n+1}(\pi_{1})$ are the intersection of the hyperplane
  $\pi_{1}:\xi_{2}=\bar{\xi}_{2}^{(1)}$ and the two-sheeted surface
  $\hat{\epsilon}_{n,n+1}$.  The projections of $\hat{C}_{n}(\pi_{1})$
  and $\hat{C}_{n+1}(\pi_{1})$ on the planes $(Im {\cal E}, \xi_{1})$
  and (Re${\cal E},\xi_{1})$ show a crossing of widths and anticrossing of
  energies, respectively. The projections of $\hat{C}_{n}(\pi_{1})$
  and $\hat{C}_{n+1}(\pi_{1})$ on the plane $(Re {\cal E}, Im {\cal
    E})$ are the trajectories of the $S-$matrix poles in the complex
  energy plane. In the figure, $\xi_{1} = d-d^{*}$. }
\end{center}
\end{figure}
\subsection{Sections of the energy surfaces}
The experimentally determined dependence of the difference of complex
resonance energy eigenvalues on one control parameter, say $\xi_{1}$,
while the other is kept constant, $\xi_{2} = \bar{\xi}^{(i)}_{2}$, i =
1,2,3,
\begin{eqnarray}\label{cuarentaiuno}
\hat{\cal E}_{1}(\xi_{1},\bar{\xi}^{(i)}_{2}) - \hat{\cal
  E}_{2}(\xi_{1},\bar{\xi}_{2}^{(i)}) =
\hat{\epsilon}_{1,2}(\xi_{1},\bar{\xi}^{(i)}_{2})
\end{eqnarray}
has a simple and straightforward geometrical interpretation, it is a
direct measurement of the intersection of the eigenenergy surface of
the doublet $\hat{\epsilon}_{1,2}(\xi_{1},\xi_{2})$ with the
hyperplane defined by the condition $\xi_{2} = \bar{\xi}_{2}^{(i)}$, i=1,2,3.
\begin{figure}
\begin{center}
\includegraphics[width=220pt,height=230pt]{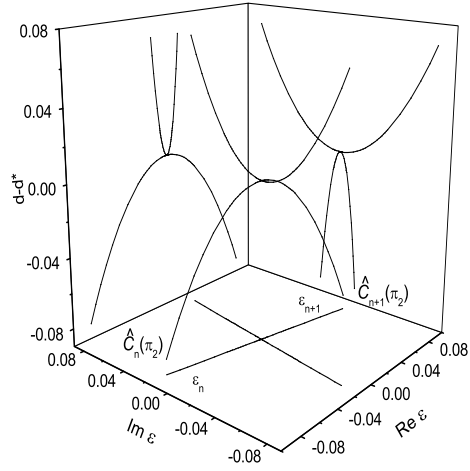}
\caption{The curves $\hat{C}_{n}(\pi_{2})$ and $\hat{C}_{n+1}(\pi_{2})$ are
  the intersections of the hyperplane $\pi_{2}$ that goes through the
  exceptional point $(\xi_{1}^{*},\xi^{*}_{2})$ in parameter space and
  the two-sheeted surface $\hat{\epsilon}_{1,2}(\xi_{1},\xi_{2})$.
  The projections of $\hat{C}_{n}(\pi_{2})$ and $\hat{C}_{n+1}
  (\pi_{2})$ on the planes $(Re {\cal E},\xi_{1})$ and $(Im {\cal
    E},\xi_{1})$ show a joint crossing of energies and widths. The
  projections of $\hat{C}_{n}(\pi_{2})$ and $\hat{C}_{n+1}(\pi_{2})$
  on the plane $(Re {\cal E}, Im {\cal E})$ are the critical
  trajectories of the $S-$matrix poles in the complex energy plane. At
  the crossing point, the two simple poles coalesce into one double
  pole of $S(E)$. }
\end{center}
\end{figure}
\begin{figure}
\begin{center}
  \includegraphics[width=220pt,height=230pt]{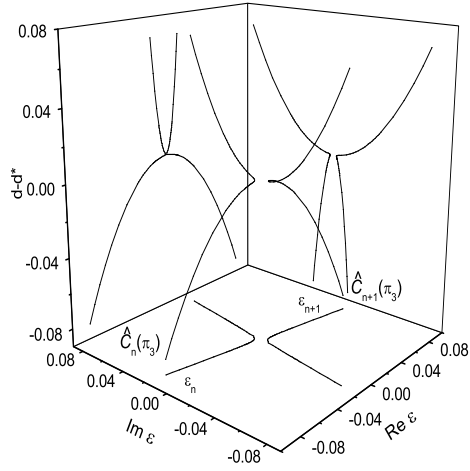}
\caption{The curves $\hat{C}_{n}(\pi_{3})$ and $\hat{C}_{n+1}(\pi_{3})
  $ are the intersection of the hiperplane $\pi_{3}:=
  \bar{\xi}_{2}^{(3)}$ and the two sheeted surface
  $\hat{\epsilon}_{n,n+1}$. The projections of $\hat{C}_{n}(\pi_{3})$
  and $\hat{C}_{n+1}(\pi_{3})$ on the planes (Re ${\cal E},\xi_{1}$)
  and (Im ${\cal E},\xi_{1}$) show a crossing of energies and an
  anticrossing of widths.  The projections of $\hat{C}_{n}(\pi_{3})$
  and $\hat{C}_{n+1}(\pi_{3})$ on the plane $( Re {\cal E}, Im {\cal
    E})$ are the trajectories of the $S-$matrix poles in the complex
  energy plane and they do not cross.  In the figure, $\xi_{1} = d -
  d^{*}$.}
\end{center}
\end{figure}

The intersection of the eigenenergy surface
$\hat{\epsilon}_{1,2}(\xi_{1},\xi_{2})$ and each one of  the hyperplanes
$\pi_{i}:\xi_{2}=\bar{\xi}_{2}^{(i)}$, defines two three-dimensional
curves for each value of $\bar{\xi}_{2}^{(i)}$
\begin{eqnarray}\label{cuarentaidos}
\hat{\epsilon}_{1,2} (\xi_{1},\xi_{2})\cap \pi_{i} = 
\cases{
\hat{C}_{1}(\pi_{i}) \\
\hat{C}_{2}(\pi_{i})} 
\end{eqnarray}
The sections $\hat{C}_{1}(\pi_{i})$ and $\hat{C}_{2}(\pi_{i})$ are the
three-dimensional curves traced by the points $\hat{\cal
  E}_{1}(\xi_{1},\bar{\xi}_{2}^{(i)})$ and $\hat{\cal
  E}_{2}(\xi_{1},\bar{\xi}_{2}^{(i)})$ on the surface
$\hat{\epsilon}_{1,2}(\xi_{1},\xi_{2})$ when the point with
coordinates $(\xi_{1},\bar{\xi}_{2}^{(i)})$ moves along a straigth
line path parallel to the $O\xi_{1}$ axis,  and $\xi_{1,i}\leq \xi_{1} \leq
\xi_{1,f}$, in parameter space.

The projections of the sections $\hat{C}_{1}(\pi_{i})$ and $\hat{C}_{2}(\pi_{i})$
on the planes (Re $\hat{\cal E}, \xi_{1}$) and (Im$\hat{\cal E}, \xi_{1}$) are
\begin{eqnarray}\label{cuarentaitres}
Re[\hat{C}_{m}(\pi_{i})] = Re \hat{\cal
  E}_{m}(\xi_{1},\bar{\xi}_{2}^{(i)}) \hspace{1cm} m = 1, 2
\end{eqnarray}
 and
\begin{eqnarray}\label{cuarentaicuatro}
Im[\hat{C}_{m}(\pi_{i})] = Im \hat{\cal
  E}_{m}(\xi_{1},\bar{\xi}^{(i)}_{2}) \hspace{1cm} m = 1, 2
\end{eqnarray}
respectively, see figures 4-6. A comparison of the representations of
the eigenenergy surfaces provided by the numerically exact computation
and the contact equivalent approximant is shown in figure 7.

The projections of the sections $\hat{C}_{1}(\pi_{i})$ and
$\hat{C}_{2}(\pi_{i})$ on the plane $(Re \hat{\cal E}$, Im
$\hat{\cal E})$ are the trajectories of the $S-$matrix poles
in the complex energy plane. An equation for these trajectories is
obtained by eliminating $\xi_{1}$ between Re$\hat{\cal
  E}_{m}(\xi_{1},\bar{\xi}_{2}^{(i)})$ and Im$\hat{\cal
  E}_{m}(\xi_{1},\bar{\xi}_{2}^{(i)}), m=1,2,$
equations(\ref{treintaisiete}-\ref{treintaiocho}),
\begin{eqnarray}\label{cuarentaicinco}
\Bigl(Re \hat{\cal E}_{m}\Bigr)^{2} - 2\cot\phi_{1}\Bigl(Re\hat{\cal
  E}_{m}\Bigr)\Bigl(Im\hat{\cal E}_{m}\Bigr) - \Bigl(Im\hat{\cal
  E}_{m}\Bigr)^{2} -
\frac{1}{4}\Bigl(\vec{R}\cdot\bar{\xi}_{c}^{(i)}\Bigr) = 0 
\end{eqnarray}
where
\begin{eqnarray}\label{cuarentaiseis}
\cot\phi_{1} = \frac{R_{2}}{I_{1}}
\end{eqnarray}
and the constant vector $\vec{\xi}^{(i)}_{c}$ is such that
\begin{eqnarray}\label{cuarentaisiete}
\Bigl(\vec{I}\cdot\vec{\xi}_{c}\Bigr)\Bigr|_{\xi_{2}=\vec{\xi}^{(i)}_{2}}
= 0
\end{eqnarray}
The discriminant of (\ref{cuarentaicinco}), $4\cot^{2}\phi_{1} + 1$, is
positive. Therefore, close to the crossing point, the trajectories of
the $S-$matrix poles are the branches of a hyperbola defined by
(\ref{cuarentaicinco}).

\begin{figure}
\begin{center}
  \includegraphics[width=300pt,height=450pt]{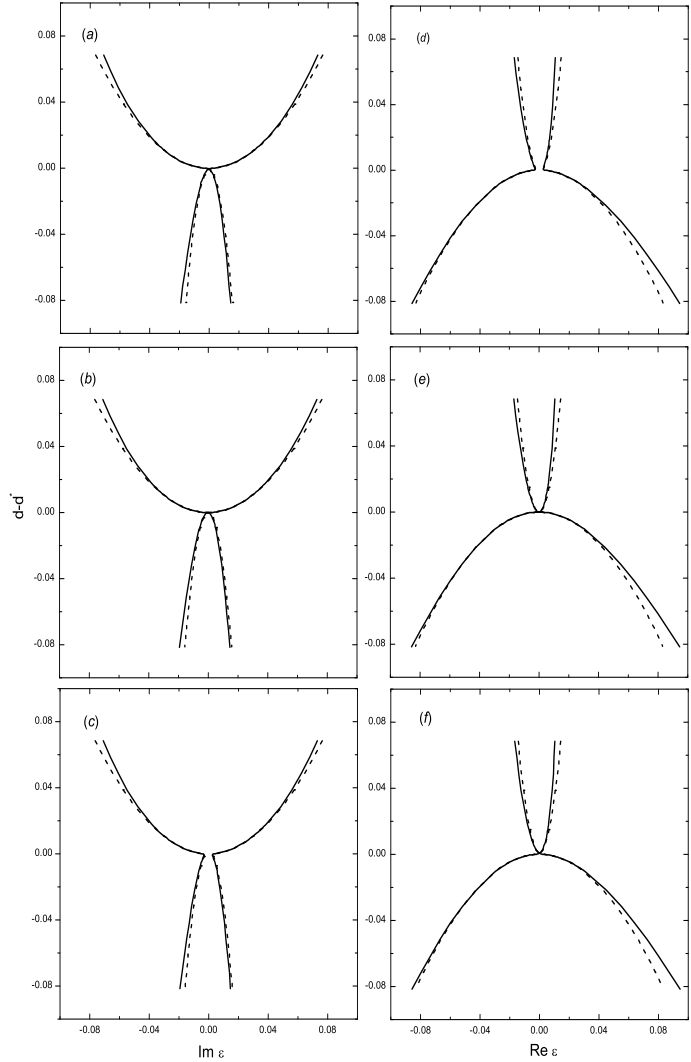}
\caption{Projections of the sections $\hat{C}_{n}(\pi_{i})$ and
  $\hat{C}_{n+1}(\pi_{i})$ on the planes (Im ${\cal E}, \xi_{1}$)
  and (Re ${\cal E},\xi_{1}$) are shown in the left and right
  columns respectively for a) $V_{3}=$ 1.0381, b) $V_{3} = V^{*}_{3}$
  and c) $V_{3}= $1.0834. The full line is the numerically exact
  calculation, the dotted line is the contact equivalent approximant. }
\end{center}
\end{figure}

\subsection{Crossings and anticrossings of energies and widths}
Crossings or anticrossings of energies and widths are experimentally
observed when the difference of complex resonance energy eigenvalues,
$\hat{\cal E}_{1}(\xi_{1},\bar{\xi}_{2}) -\hat{\cal
  E}_{2}(\xi_{1},\bar{\xi}_{2}) = \Delta E - i 1/2\Delta\Gamma$ is
measured as a function of the slowly varying parameter $\xi_{1}$,
keeping the other constant, $\xi_{2}=\bar{\xi}_{2}^{(i)}$. 
From equations (\ref{treintaisiete}-\ref{treintaiocho}), and keeping
$\xi_{2}=\bar{\xi}^{(i)}_{2}$, we obtain
\begin{equation}\label{cuarentaiocho}
\begin{array}{c}
\Delta E = E_{n}-E_{n+1} 
= \frac{\sqrt{2}}{2}\Bigl[{}_{+}
\sqrt{(\vec{R}\cdot\vec{\xi})^{2}+
(\vec{I}\cdot\vec{\xi})^{2}}
+ (\vec{R}\cdot\vec{\xi})\Bigr]^{1/2}\Bigr]_{\xi_{2}=\bar{\xi}_{2}^{(i)}}
\end{array}
\end{equation}
and
\begin{eqnarray}\label{cuarentainueve}
\Delta\Gamma = \Bigl(\Gamma_{n} - \Gamma_{n+1}\Bigr) 
= -\sqrt{2} \Bigl[{}_{+}
\sqrt{(\vec{R}\cdot\vec{\xi})^{2}+(\vec{I}\cdot\vec{\xi})^{2}} -
(\vec{R}\cdot\vec{\xi})\Bigr]^{1/2}\Bigr|_{\xi_{2}=\bar{\xi}_{2}^{(i)}}\cr
\end{eqnarray}
These expressions allow us to relate the terms $(\vec{R}\cdot\vec{\xi})$
and $(\vec{I}\cdot\vec{\xi})$ directly with observables of the isolated
doublet of resonances.
Taking the product of  $\Delta E\Delta\Gamma$, and
recalling equation (\ref{treintainueve}), we get
\begin{eqnarray}\label{cincuenta}
\Delta E\Delta\Gamma = -\Bigl(\vec{I}\cdot\vec{\xi}\Bigr)\Bigr|_{\xi_{2}=
\bar{\xi}_{2}^{(i)}}
\end{eqnarray}
and taking the differences of the squares of the left hand sides of
(\ref{cuarentaiocho}) and (\ref{cuarentainueve}), we get
\begin{eqnarray}\label{cincuentaiuno}
\Bigl(\Delta E\Bigr)^{2} - \frac{1}{4}\Bigl(\Delta\Gamma\Bigr)^{2} =
\Bigl(\vec{R}\cdot\vec{\xi}\Bigr)\Bigr|_{\xi_{2}=\bar{\xi}_{2}^{(i)}}
\end{eqnarray}

At a crossing of energies $\Delta E$ vanishes, and at a crossing of
widths $\Delta\Gamma$ vanishes.  Hence, the relation found in
eq.(\ref{cincuenta}) means that a crossing of energies or widths
can occur if and only if
$(\vec{I}\cdot\vec{\xi})_{\bar{\xi}^{(i)}_{2}}$ vanishes

For a vanishing $(\vec{I}\cdot\vec{\xi}_{c})_{\bar{\xi}^{(i)}_{2}} = 0
= \Delta E\Delta\Gamma$, we find three cases, which are distinguished
by the sign of $(\vec{R}\cdot\vec{\xi}_{c})_{\bar{\xi}^{(i)}_{2}}$.
From eqs. (\ref{cuarentaiocho}) and (\ref{cuarentainueve}),
\begin{enumerate}
\item {\it $(\vec{R}\cdot\vec{\xi}_{c})_{\bar{\xi}^{(i)}_{2}} > 0 $
implies $\Delta E \neq 0$ and $\Delta\Gamma = 0$, that is, energy
anticrossing and width  crossing}.
  \item {\it $(\vec{R}\cdot\vec{\xi}_{c})_{\bar{\xi}^{(i)}_{2}} = 0 $
      implies $\Delta E = 0$ and $\Delta\Gamma = 0$, that is, joint
      energy and width crossings, which is also degeneracy of the two
      complex resonance energy eigenvalues}.
    \item {\it $(\vec{R}\cdot\vec{\xi}_{c})_{\bar{\xi}^{(i)}_{2}} < 0
        $ implies $\Delta E = 0$ and $\Delta\Gamma\neq 0$, i.e. energy
        crossing and width anticrossing}.
\end{enumerate}

This rich physical scenario of crossings and anticrossings for the
energies and widths of the complex resonance energy eigenvalues
extends a theorem of von Neumann and Wigner \cite{five} for bound
states to the case of unbound states.

The general character of the crossing-anticrossing relations of the
energies and widths of a mixing isolated doublet of resonances,
discussed above, has been experimentally established by P. von
Brentano and his collaborators in a series of beautiful experiments
\cite{thirtyseven,thirtyeight,thirtynine}.

\subsection{Trajectories of the $S-$matrix poles and changes of
  identity}
In subsection 4.1, we found that, when one control parameter, say
$\xi_{1}$, is varied and the other control parameter is kept constant
$\bar{\xi}^{(i)}_{2}$ and close to the exceptional value, the
trajectories of the $S-$matrix poles are the branches of a hyperbola
defined by (\ref{cuarentaicinco}-\ref{cuarentaisiete}). The
asymptotes of this hyperbola are the two straight lines defined by
\begin{eqnarray}\label{cincuentaidos}
Im {\cal E}^{(I)} = \tan \frac{\phi}{2}Re {\cal E}^{(I)}
\end{eqnarray}
and
\begin{eqnarray}\label{cincuentaitres}
Im {\cal E}^{(II)} = - \cot\frac{\phi}{2}Re {\cal E}^{(II)}
\end{eqnarray}
The two branches of the hyperbola are in opposite quadrants of the
complex energy plane divided by the asymptotes, see figure 8.
\begin{figure}
\begin{center}
  \includegraphics[width=300pt,height=450pt]{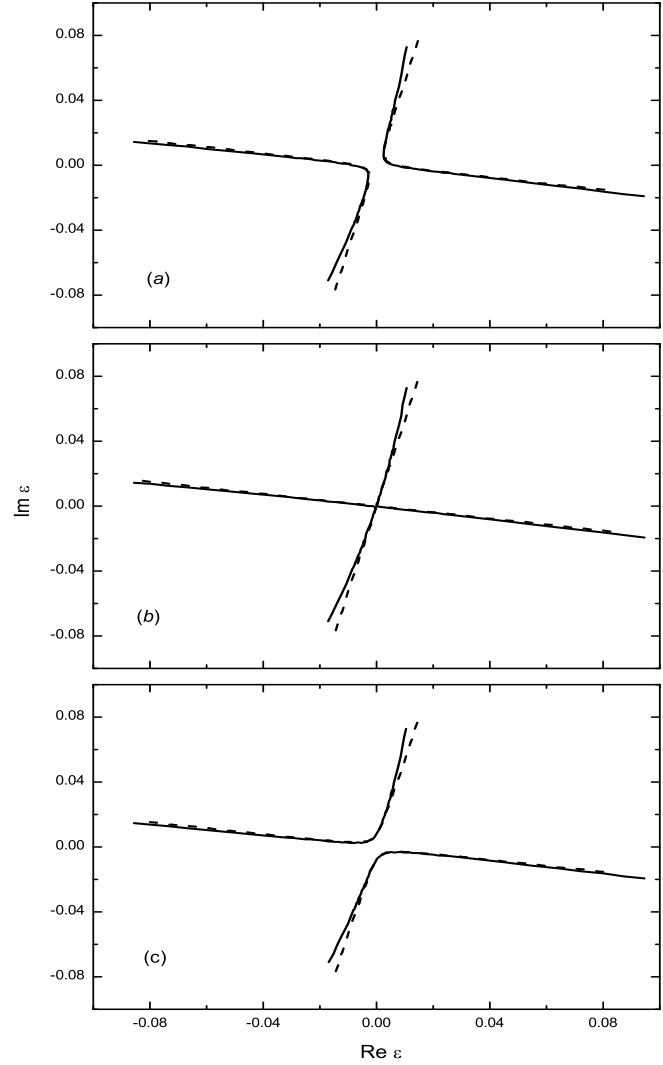}
  \caption{Trajectories of the poles of the scattering matrix,
    $S({\cal E})$, of an isolated doublet of resonances in a double
    barrier potential close to a degeneracy of unbound states. The
    control parameters are the width, $d$, of the inner barrier and the
    depth, $V_{3}$, of the outer well. The trajectories are traced by
    the poles ${\cal E}_{1}(d,V^{(i)}_{3})$ and ${\cal
      E}_{2}(d,V^{(i)}_{3})$ on the complex ${\cal E}-$plane when the
    point $(d,V^{(i)}_{3})$ moves on the straight line path $\pi_{i}$;
    $V_{3}=V^{(i)}_{3}$. The top, middle, and bottom figures show the
    trajectories corresponding to
    $(\vec{R}\cdot\vec{\xi}_{c})_{\bar{\xi}^{(i)}_{2}} > 0$,
    $(\vec{R}\cdot\vec{\xi}_{c})_{\bar{\xi}^{(i)}_{2}}  = 0$, and
    $(\vec{R}\cdot\vec{\xi}_{c})_{\bar{\xi}^{(i)}_{2}} < 0$, respectively,
    with $(\xi_{1},\xi_{2}) = (d-d^{*}, V_{3}-V^{*}_{3})$.
    The full line is the numerically exact calculation, the dotted
    line is the contact equivalent approximant. }
\end{center}
\end{figure}

We find three types of trajectories, which are distinguished by the
sign of $(\vec{R}\cdot\vec{\xi})|_{\xi_{2}=\bar{\xi}_{2}}$:
\begin{enumerate}
\item When $(\vec{R}\cdot\vec{\xi})|_{\xi_{2}=\bar{\xi}_{2}}> 0$, one
  branch of the hyperbola lies to the left and the other lies to the
  right of a vertical line that goes through the crossing point.
  
\item Critical trajectories, when
  $(\vec{R}\cdot\vec{\xi})|_{\xi_{2}=\bar{\xi}_{2}}= 0 $, the
  trajectories are the asymptotes of the hyperbola. The two simple
  poles start from opposite ends of the same straight line and move
  towards each other until they meet at the crossing point where they
  coalesce to form a double pole of the $S-$matrix. From there, they
  separate moving away from each other on a straight line at
  $90^{\circ}$ with respect to the first asymptote.

\item When $(\vec{R}\cdot\vec{\xi})|_{\xi_{2}=\bar{\xi}_{2}} < 0 $,
  one pole moves on one branch of the hyperbola that lies above and
  the other pole moves on the other branch that lies below a
  horizontal straight line that goes through the crossing point.

\end{enumerate}

\begin{figure}
\begin{center}
  \includegraphics[width=435pt,height=295pt]{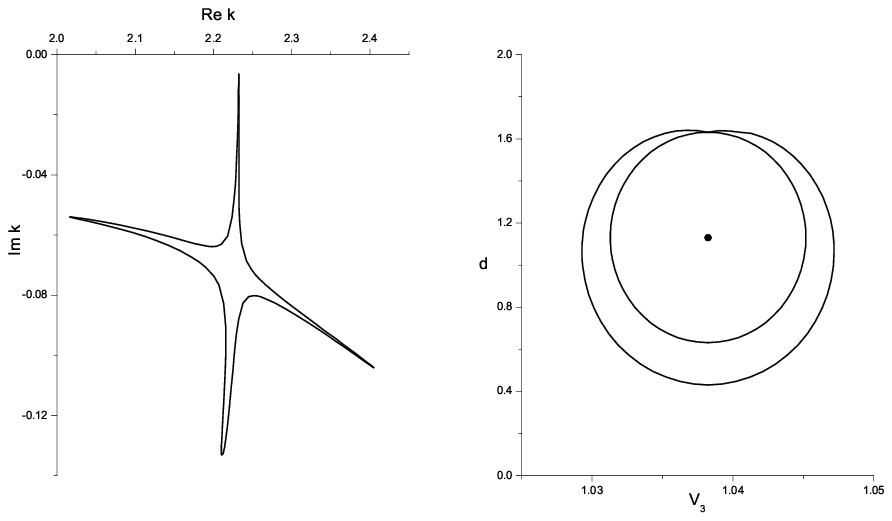}
  \caption{The two poles of the isolated doublet of unbound states
    trace a starlike trajectory in the complex $k-$plane, shown on the
    left hand side of the figure, when the physical system is
    transported in parameter space twice around the exceptional point
    in the circular path shown at the right hand side of the figure.
    For an explanation, see the text.}
\end{center}
\end{figure}

When a small change in the control parameter $\bar{\xi}_{2}^{(i)}$
changes the sign of $(\vec{R}\cdot\vec{\xi})|_{\xi_{2}=\bar{\xi}_{2}}$, it
produces a small change in the initial position of the poles, but the
trajectories change suddenly from type (i) to type (iii), this very
large and sudden change of the trajectories exchanges almost exactly
the final position of the poles as can be appreciated from figure 8.
This dramatic change was called a ``change of identity'' by Vanroose,
van Leuven, Arickx and Broeckhove \cite{eight} who discussed an
example of this phenomenon in the $S-$matrix poles in a two-channel
model, Vanroose \cite{fourty} has also discussed these properties in
the case of the scattering of a beam of particles by a double barrier
potential with two regions of trapping.

\subsection{Changes of identity when going around the exceptional point}

In the previous discussion of the trajectories of the $S-$matrix poles
and their changes of identity, the two straight lines, parallel to the
$O\xi_{1}$ axis, defined by the conditions $\xi_{1,i} \leq \xi_{1} \leq \xi_{1,f}$
and $\xi_{2} = \bar{\xi}^{(i)}_{2}$, i = 1,3, are the two long sides of a
very long and narrow rectangle in parameter space that surrounds the
exceptional point. The two short sides of this rectangle are defined
by the small change in $\xi_{2}$ when going from $\bar{\xi}^{(1)}_{2}$ to
$\bar{\xi}^{(3)}_{2}$, keeping $\xi_{1,i}$ or $\xi_{1,f}$ constant, to which
we refered above, when explaining the changes of identity of the poles
of the doublet of unbound states. In order to understand better the
meaning of the changes of identity of those poles, it will be
convenient to deform continuously the long and narrow rectangle into a
circle and consider the motion of the zeroes of the Jost function
$k_{1}(x_{1},x_{2})$ and $k_{2}(x_{1},x_{2})$ in the complex
$k-$plane, when the system is transported in parameter space around
the exceptional point in the closed circular path  equivalent by
continuous deformation to the long and narrow rectangle of the
previous discussion.

On the left hand side of figure 9, we show the trajectory traced in
the complex $k-$plane by the two zeroes of the Jost
function, $k_{1}(x_{1},x_{2})$ and $k_{2}(x_{1},x_{2})$, when the
physical system is transported in parameter space twice around the
exceptional point on the double circular circuit shown on the right
hand side of the same figure. When the system is at the crossing point
of the two circular paths in parameter space, one zero, say
$k_{1}(x_{1},x_{2})$, is at the uppermost point and the other, say
$k_{2}(x_{1},x_{2})$, is at the lowermost point of the four points
star in the complex $k-$plane.
\begin{enumerate}
\item As the point representing the system in parameter space moves on
  the inner circle in counterclockwise direction, from its initial
  position until it makes one complete round about the exceptional
  point and is back at its initial position, the $k_{1}(x_{1},x_{2})$
  zero moves on the star in the complex $k-$plane in clockwise
  direction, from its initial position at the topmost point of the
  star, goes through the point at the extreme right hand side of the
  star and ends at the lowest point in the star, while the
  $k_{2}(x_{1},x_{2})$ zero moves also in clockwise direction, from
  its initial position at the lowest point in the star goes trough the
  point in the extreme left hand side of the star and ends at topmost
  point on the star. It follows that, when the system goes around the
  exceptional point once in parameter space, the poles of the
  scattering matrix are exactly exchanged.
\item As the point representing the system in parameter space
  continues its conterclockwise motion, now on the outercircle, until
  it completes a second round about the exceptional point and is back
  at its initial position, the $k_{1}(x_{1},x_{2})$ zero moves on the
  star in the complex $k-$plane in clockwise direction from the lowest
  point on the star, goes through the extreme left hand side point on
  the star and ends at the topmost point on the star, while the
  $k_{2}(x_{1},x_{2})$ zero moves on the star in the complex $k-$plane
  also in clockwise direction, from the topmost point on the star,
  goes through the extreme right hand side point on the star and ends
  at lowest point of the four points star. Therefore, when the system goes
  around the exceptional point twice in parameter space the poles of
  the $S(k)$ matrix and the complex energy eigenvalues return to their
  initial values in the complex $k-$plane. The eigenfunctions also
  return to their initial values but they acquire a geometric phase
  \cite{nineteen,twenty,twentyone,twentytwo,alexei,berry0}.
\end{enumerate}

\section{Summary and conclusion}
In this paper, we discussed some mathematical and physical aspects of
the non-Hermitian degeneracy of two unbound energy eigenstates of a
Hamiltonian dependending on two control parameters. We solved
numerically the implicit transcendental equation that defines the
eigenenergy surface of a degenerating isolated doublet of unbound
states in the simple but illustrative case of the scattering of a beam
of particles by a double square barrier potential. The analytical
characterization of the singularity of the energy surface was made in
the more general case of a short ranged potential with two regions of
trapping. We showed that, from the explicit knowledge of the Jost
function as a function of the control parameters of the system, it is
possible to derive a two parameter family of functions which is
contact equivalent to the exact energy-pole position function at the
degeneracy point and includes all small perturbations of the
degeneracy conditions. This unfolding of the degeneracy point gives a
simple and explicit, but very accurate, representation of the
eigenenergy surface close to the exceptional point, see figure 7. In
parameter space, the surface that represents the complex energy
eigenvalues has a branch point of square root type at the crossing
point, and branch cuts in its real and imaginary parts that start at
the exceptional point but extend in opposite directions in parameter
space. In the complex energy plane, the crossing point of two simple
resonance poles of the scattering matrix is an isolated point where
the scattering matrix has one double resonance pole. Crossings and
anticrossings of the energies and widths of the resonances in an
isolated doublet of unbound states of a quantum system, as well as the
sudden change in the shape of the $S-$matrix pole trajectories,
observed when one control parameter is varied while the other is kept
constant at a value close to the exceptional value, are fully explained
in terms of sections of the energy surfaces.

\section{Acknowledgements}
 We thank Professor Peter von Brentano (Universit\"at zu K\"oln) for
 many inspiring discussions on this exciting problem. 
 This work was partially supported by CONACyT M\'exico under Contract
 No. 40162-F and by DGAPA-UNAM Contract No. PAPIIT: \\ IN116202


\section*{References}
\begin{thebibliography}{10}
\bibitem{one} Gamow G 1928 {\it Z. Phys} {\bf 51} 204
\bibitem{two} Peierls R E 1959 {Proc. Roy. Soc. (London) Ser. A}  {\bf
    253} 16
\bibitem{three}Hern\'andez E, J\'auregui A,  and Mondrag\'on
  A 2003 {\it Phys. Rev. A} {\bf 67} 022721
\bibitem{four}Berry M V 2004 {\it Czech. J. Phys.} {\bf 54} 1039
\bibitem{five}von Neumann J and Wigner E P 1929 {\it Physik Z.}
  {\bf 30} 467
\bibitem{six}Teller E 1937 {\it J. Phys. Chem.} {\bf 41} 109
\bibitem{seven}von Brentano P 1990 {\it Phys. Lett. B } {\bf 238}
  1; 1990 {\it Phys. Lett. B} {\bf 246} 320; 1991 {\it Phys. Lett. B}
  {\bf 265}14
\bibitem{eight}Vanroose W, Leuven P, Arickx F, and Broeckhove
  J 1997 {\it J. Phys. A: Math. Gen.} {\bf 30} 5543
\bibitem{nine}Friedrich H and Wintgen D 1985 {\it Phys. Rev. A} {\bf
    32} 3231
\bibitem{ten}Mondrag\'on A, and Hern\'andez E 1993 {\it J. Phys. A:
      Math.  Gen.} {\bf 26} 5595

\bibitem{eleven}Hern\'andez E and Mondrag\'on A (1994) {\it Phys. Lett. B}
  {\bf 326} 1
\bibitem{twelve}von Brentano P 1996 {\it Phys. Rep.} {\bf 264} 57
\bibitem{thirten} Antoniou I, Gadella M and Pronko G 1998 {\it
    J. Math. Phys.} {\bf 39} 2429
\bibitem{fourteen} Bohm A, Loewe M, Maxson S, Patuleanu, P\"untmann C
  and Gadella M 1997 {\it J. Math. Phys.} {\bf 38} 6072
\bibitem{fifteen}Lassila K E and Ruuskanen V 1966 {\it
    Phys. Rev. Lett.} {\bf 17} 490
\bibitem{sixteen} Knight P L 1979 {\it Phys. Lett. A} {\bf 72} 309
\bibitem{seventeen}Kylstra N J and Joachain C J 1996 {\it
    Europhys. Lett.} {\bf 36} 657
\bibitem{eighteen}Kylstra N J and Joachain C J 1998 {\it Phys. Rev. A}
  {\bf 57} 412
\bibitem{znojil}Znojil M 2006 {\it J. Phys A: Math. Gen.} {\bf 39} 441
\bibitem{znojil1}Znojil M 1999 {\it Phys. Lett. A} {\bf 259} 220
\bibitem{znojil2}Znojil M Levai G, 2001 \emph{Mod. Phys. Lett.} A 
\textbf{16} 2273
\bibitem{mostafa}Mostafazadeh A 2002 \emph{J. Math. Phys.} \textbf{43} 6443

\bibitem{rotter1}Rotter I, Sadreev A F 2005 {\it Phys. Rev. E} {\bf 71} 
036227
\bibitem{rotter2}Rotter I 2003 {\it Phys. Rev. E} {\bf 67} 026204
\bibitem{rotter3}Rotter I 2002 {\it Phys. Rev. E} {\bf 65} 026217
\bibitem{rotter4}Magunov A L, Rotter I, Strakhova S I 2001 {\it J.
    Phys B: Atom. Mol. Opt. Phys.} {\bf 34} 29

\bibitem{nineteen}Hern\'andez E, J\'auregui A, and Mondrag\'on
 A 1992 {\it Rev. Mex. Fis.} {\bf 38}, {\it Suppl 2},  128
\bibitem{twenty}Mondrag\'on A and Hern\'andez E 1996 {\it J. Phys. A:
    Math.  Gen.} {\bf 29} 2567
\bibitem{twentyone}Mondrag\'on A and Hern\'andez E 1998
  Accidental degeneracy and Berry phase of resonant states 
  {\it Irreversibility and Causality: Semigroups and Rigged Hilbert
    Sapce (Lecture Notes in Physics) vol 504 } Ed A. Bohm, H-D Doebner
  and P Kielanowski (Berlin: Springer-Verlag) p 257
\bibitem{twentytwo}Heiss W D 1999 {\it Eur. Phys. J.} D {\bf 7}  1
\bibitem{alexei}Mailybaev Alexei A, Kirillov Oleg N and Seyranian
  Alexander P 2005 {\it Phys. Rev. A} {\bf 72} 014104
\bibitem{twentythree} Dembowski C, Gr\"af H D, Harney H L, Heine
  A, Heiss W D, Rehfeld H  and Richter A 2001 {\it Phys. Rev. Lett.}
  {\bf 86} 787
\bibitem{twentyfour}Dembowski C, Dietz B, Gr\"af H D, Harney
  H L, Heine A, Heiss  W D, and Richter  A 2003 {\it Phys. Rev.
    Lett}. {\bf 90} 034101
\bibitem{twentyfive}Hern\'andez  E, J\'auregui A,  and Mondrag\'on
  A 2005 {\it Phys. Rev. E} {\bf 72} 026221
\bibitem{twentysix}Berry M V and Dennis M R 2003 {\it Proc. R. Soc. Lond. 
  A} {\bf 459} 1261
\bibitem{twentyseven}Keck F, Korsch H J, and Mossmann S 2003
  {\it J. Phys.  A: Math.  Gen.} {\bf 36} 2125

\bibitem{twentyeight}Korsch H J, and Mossmann S 2003 {\it J. Phys. A:
    Math.  Gen.} {\bf 36} 2139

\bibitem{twentynine}Shuvalov A L and Scott N 2000 {\it Acta Mech.} {\bf
    140} 1
\bibitem{thirty} Seyranian A P, Kirillov O N and Mailybaev A A 2005
  {\it J. Phys. A: Math. Gen.} {\bf 38} 1723
\bibitem{thirtyone} Kirillov O N, Mailybaev A A and Seyranian A P 2005
  {\it J. Phys. A: Math. Gen.} {\bf 38} 5531
\bibitem{heiss2}Stehmann T, Heiss W D, Scholtz F G 2004 {\it J. Phys.
      A: Math. Gen.} {\bf 37} 7813
\bibitem{heiss3}Heiss W D, 2004 {\it J. Phys. A: Math. Gen.} {\bf 37} 2455
\bibitem{thirtytwo}Newton R G 1982 {\it Scattering Theory of Waves and
    Particles}, 2nd. edn. (Berlin: Springer-Verlag) Chapt. 12
\bibitem{thirtythree}Boas R P 1954 {\it Entire Functions} (Academic New
  York) p. 22
\bibitem{thirtyfour}Pfluger A 1943 {\it Communs. Math. Helv.} {\bf 16} 1 
\bibitem{thirtyfive}Krantz S G  and Parks H R  2002 {\it The Implicit
    Function Theorem} (Boston:Birkh{\"a}user) Chapt. 5
\bibitem{thirtysix}Seydel R 1991 {\it Practical Bifurcation and Stability
    Analysis. IAM5} 2nd. edn. (New York: Springer-Verlag)
  Chapt. 8
\bibitem{thirtyseven}von Brentano P, and Philipp M 1999 {\it
    Phys. Lett. B} {\bf 454} 171
\bibitem{thirtyeight}Philipp M, von Brentano P, Pascovici G and
  Richter A 2000 {\it Phys. Rev. E} {\bf 62}  1922
\bibitem{thirtynine}von Brentano P 2000 {\it Rev. Mex. Fis}{\bf 48},
  {Suppl 2}, 1
\bibitem{fourty}Vanroose W 2001 {\it Phys. Rev. A} {\bf 64} 062708
\bibitem{berry0}Berry M V 1984 \emph{Proc. R. Soc. } (London) A 
\textbf{392} 45
\end{thebibliography}
\end{document}